\documentclass[12pt]{article}  %fuer beidseitige Version
\usepackage[a4paper,textwidth=490.0pt,textheight=703.1pt]{geometry}
\usepackage{cite}
\usepackage{graphicx}
\usepackage{epsfig}
\usepackage{amsmath}
\usepackage{amssymb}
\usepackage{ulem}
\usepackage{mdwlist}
\usepackage{color}
\usepackage{float}
\usepackage[rflt]{floatflt}
\usepackage{slashed}
\usepackage{array}

\usepackage[english]{babel}
\usepackage[latin1]{inputenc}
\usepackage[T1]{fontenc}
\usepackage{ae}

\usepackage{url}

\usepackage{amsmath, amsthm, amssymb}

\newtheorem{thm}{Theorem}[section]

\usepackage{array}

\usepackage[english]{babel}
\usepackage[latin1]{inputenc}
\usepackage[T1]{fontenc}
\usepackage{ae}

\newtheorem{definition}[thm]{Definition}

\newcommand{\gsim}{\raisebox{-0.07cm   }
{$\, \stackrel{>}{{\scriptstyle\sim}}\, $}}
\newcommand{\GeV}{\rm GeV}

\newcommand{\Li}{{\rm Li}}
\newcommand{\Mvec}{{\rm \bf M}}

\newcommand{\ep}{\varepsilon}

\usepackage{rotating}

\usepackage{graphicx}
\newcounter{mmacnt}
\def\restartmma{\setcounter{mmacnt}{0}}
\restartmma \catcode`|=\active
\def|#1|{\mathrm{#1}}
\catcode`|=12
\newenvironment{mma}{
 \par\smallskip
 \catcode`|=\active
 \parskip=0pt\parindent=0pt % locally
 \small
 \def\In##1\\{%
   \def\linebreak{\hfill\break\null\qquad}%
   \refstepcounter{mmacnt}
   \hangindent=2.5em\hangafter=0
   \leavevmode
   \llap{\tiny\sffamily In[\arabic{mmacnt}]:=\kern.5em}%
   \mathversion{bold}\footnotesize$\displaystyle##1$\normalsize
   \mathversion{normal}\par
 }%
 \def\Print##1\\{%
   \def\linebreak{\hfill\break}%
   \hangindent=2.5em\hangafter=0
   \leavevmode ##1\par}%
 \def\Out##1\\{%
   \def\linebreak{$\hfill\break\null\hfill$}%
   \kern\abovedisplayskip\par
   \hangindent=2.5em\hangafter=0
   \leavevmode
   \llap{\tiny\sffamily Out[\arabic{mmacnt}]=\kern.5em}
   \footnotesize$\displaystyle##1$\normalsize\hfill\null\par
   \kern\belowdisplayskip
 }%
 \def\Warning##1##2\\{%
   \def\linebreak{\hfill\break}%
   \hangindent=2.5em\hangafter=0
   \leavevmode
   {\scriptsize##1 : ##2}\par}%
}{%
 \par\smallskip
}

\usepackage{color}

\newenvironment{fshaded}{%
\MakeFramed {\FrameRestore}
}%
{\endMakeFramed}

\usepackage{tikz}
\usetikzlibrary{matrix}

\allowdisplaybreaks[4]

\begin{document}
\setlength{\baselineskip}{0.515cm}
\sloppy
\thispagestyle{empty}
\begin{flushleft}
DESY 15--077
\hfill %%{\tt arXiv:1505.xxxxx [hep-ph]}
\\
DO--TH 15/07\\
MITP/15-039\\
August 2015\\
\end{flushleft}

\mbox{}
\vspace*{\fill}
\begin{center}

{\LARGE\bf The \boldmath $O(\alpha_s^3)$ Heavy Flavor Contributions to the}

\vspace*{2mm}
{\LARGE\bf Charged Current \boldmath Structure Function $xF_3(x,Q^2)$}

\vspace*{2mm}
{\LARGE\bf  at Large Momentum Transfer}

\vspace{4cm}
\large
A.~Behring$^a$,
J.~Bl\"umlein$^a$,
A.~De Freitas$^{a,b}$, 
A.~Hasselhuhn$^c$,

\vspace*{1.5mm}
A.~von~Manteuffel$^d$,
and
C.~Schneider$^b$

\vspace{1.5cm}
\normalsize   
{\it  $^a$~Deutsches Elektronen--Synchrotron, DESY,}\\
{\it  Platanenallee 6, D-15738 Zeuthen, Germany}

\vspace*{3mm}
{\it $^b$~Research Institute for Symbolic Computation (RISC),\\
                          Johannes Kepler University, Altenbergerstra\ss{}e 69,
                          A--4040, Linz, Austria}\\

\vspace*{3mm}   
{\it  $^c$ 
Institute for Theoretical Particle Physics,
Karlsruhe Institute of Technology (KIT)}\\
{\it D-76128 Karlsruhe},

\vspace*{3mm}   
{\it  $^d$ PRISMA Cluster of Excellence,  Institute of Physics, J. Gutenberg
University,}\\
{\it D-55099 Mainz, Germany.}
\\

\end{center}
\normalsize
\vspace{\fill}
\begin{abstract}
\noindent 
We calculate the massive Wilson coefficients for the heavy flavor contributions to the non-singlet charged 
current deep-inelastic scattering structure function $xF_3^{W^+}(x,Q^2)+xF_3^{W^-}(x,Q^2)$ in the asymptotic 
region $Q^2 \gg m^2$ to 3-loop order in Quantum  Chromodynamics (QCD) at general values of the Mellin variable $N$ 
and the momentum fraction $x$. Besides the heavy quark pair production also the single heavy flavor excitation  
$s \rightarrow c$ contributes. Numerical results are presented for the charm quark contributions and consequences 
on the Gross-Llewellyn Smith sum rule are discussed.
\end{abstract}

\vspace*{\fill}
\noindent
\numberwithin{equation}{section}

\newpage
%%%%%%%%%%%%%%%%%%%%%%%%%%%%%%%%%%%%%%%%%%%%%%%%%%%%%%%%%%%%%%%%%%%%%%%%%%%%%%%%%%%%%%%%%%%%%%%%%%%
\section{Introduction}
\label{sec:1}
%%%%%%%%%%%%%%%%%%%%%%%%%%%%%%%%%%%%%%%%%%%%%%%%%%%%%%%%%%%%%%%%%%%%%%%%%%%%%%%%%%%%%%%%%%%%%%%%%%%

\vspace*{1mm}
\noindent
The non-singlet structure function $xF_3(x,Q^2)$ has been a first experimental source to determine the
valence quark distributions $u_v(x,Q^2)$ and $d_v(x,Q^2)$ of the nucleon in deep-inelastic charged current 
neutrino-nucleon scattering \cite{Eisele:1986uz,Diemoz:1986kt,Schmitz}. Flavor non-singlet distributions allow 
for clear measurements of the strong coupling constant $a_s =  \alpha_s/(4\pi) \equiv g_s^2/(4\pi)^2$ \cite{Blumlein:2006be}, 
due to the absence of gluonic effects in the QCD evolution \cite{Blumlein:1997em}. 

Massless and massive QCD corrections have been calculated in 
Refs.~\cite{Bardeen:1978yd,Furmanski:1981cw,Gluck:1997sj,Blumlein:2011zu} to first order in the strong coupling 
constant\footnote{The massive 1-loop corrections given in \cite{Gottschalk:1980rv} were corrected in \cite{Gluck:1997sj}, 
see also \cite{Blumlein:2011zu}.} and in 
Refs.~\cite{Zijlstra:1992kj,Moch:1999eb,Buza:1997mg,Blumlein:2014fqa,Hasselhuhn:2013swa}
to
$O(a_s^2)$\footnote{Some results given in \cite{Buza:1997mg} have been corrected in Ref.~\cite{Blumlein:2014fqa}.}. 
The massive $O(a_s^2)$ corrections were calculated 
in the asymptotic representation \cite{Buza:1995ie}, which is valid at high scales $Q^2$. One may 
perform an $O(a_s)$ comparison for the process of single heavy quark excitation. Here the approximation holds for
$Q^2/m^2 \gsim 50$ \cite{Alekhin:2014sya}, where $Q^2 = -q^2$ denotes the virtuality of the 
4-momentum transfer $q$, and $m$ is
the heavy quark mass.

The present data on the neutrino structure function 
$xF_3(x,Q^2)$ or similar measurements at HERA \cite{PDG} have not yet reached the level of precision of 1-2 \%, as is the 
case for the structure function $F_2(x,Q^2)$ in neutral current 
deep-inelastic scattering. However, at neutrino factories 
planned for the future \cite{Kaplan:2014xda} this situation will change and even 3-loop QCD corrections will be of importance.
The experimental precision reached for charged current interactions at the $ep$ collider HERA can 
also be further refined at future 
high energy facilities probing deep-inelastic scattering, such as LHeC \cite{AbelleiraFernandez:2012cc} and EIC 
\cite{EIC}. It is 
expected that non-singlet data taken at these facilities will help to improve further the knowledge of the strong coupling 
constant $\alpha_s(M_Z^2)$ \cite{Bethke:2011tr}. The 3-loop massless Wilson coefficient for the structure function 
combination $xF_3^{W^+}(x,Q^2)+xF_3^{W^-}(x,Q^2$ 
has been calculated in Ref.~\cite{Moch:2008fj}. In the present paper we calculate the massive 
3-loop Wilson coefficient for this combination in the asymptotic region $Q^2 \gg m^2$.

The charged current structure functions $xF_3^{W^\pm}(x,Q^2)$ may be measured both in neutrino- 
and charged lepton-nucleon scattering.
In the case of single gauge boson exchange the differential scattering cross sections read 
\cite{Arbuzov:1995id,Blumlein:2012bf}~:
%---------------------------------------------------------------------------------
\begin{eqnarray}
\frac{d \sigma^{\nu(\bar{\nu})}}{dx dy} 
&=& \frac{G_F^2 s}{4 \pi} \frac{M_W^4}{(M_W^2 + Q^2)^2} 
\\ && \times
\Biggl\{
\left(1 + (1-y)^2\right) F_2^{W^\pm}(x,Q^2)
- y^2 F_L^{W^\pm}(x,Q^2)
\pm \left(1 - (1-y)^2\right) xF_3^{W^\pm}(x,Q^2)\Biggr\}
\nonumber\\
\frac{d \sigma^{l(\bar{l})}}{dx dy} 
&=& \frac{G_F^2 s }{4 \pi} \frac{M_W^4}{(M_W^2 + Q^2)^2} 
\\ && \times
\Biggl\{
\left(1 + (1-y)^2\right) F_2^{W^\mp}(x,Q^2)
- y^2 F_L^{W^\mp}(x,Q^2)
\pm \left(1 - (1-y)^2\right) xF_3^{W^\mp}(x,Q^2)\Biggr\}~.
\nonumber
\end{eqnarray}
%---------------------------------------------------------------------------------
Here $x = Q^2/ys$ and $y = q.P/l.P$ denote the Bjorken variables, $l$ and $P$ are the incoming lepton and nucleon 4-momenta,
and $s = (l+P)^2$. $G_F$ is the Fermi constant and $M_W$ 
the mass of the $W$-boson. $F_i^{W^\pm}(x,Q^2)$ are the structure functions, where the $+(-)$ signs refer to incoming neutrinos 
(antineutrinos) and charged antileptons (leptons), respectively. We will consider the combination of structure functions
%---------------------------------------------------------------------------------
\begin{eqnarray}
xF_3^{W^+ - W^-}(x,Q^2) = xF_3^{W^+}(x,Q^2) + xF_3^{W^-}(x,Q^2)
\end{eqnarray}
%---------------------------------------------------------------------------------
in the following. It can be measured projecting onto the kinematic factor $Y_- = 1 - (1-y)^2$ of 
the differential cross 
sections at $x, Q^2 = \rm const.$ as a difference. It is given by 
%---------------------------------------------------------------------------------
\begin{eqnarray}
F_3^{W^+ - W^-}(x,Q^2) &=& 
\Bigl[
  |V_{du}|^2 (d - \overline{d}) 
+ |V_{su}|^2 (s - \overline{s}) 
+ V_{u}      (u - \overline{u})\Biggr] \otimes \Biggl[
C_{q,3}^{W^+ - W^-, \rm NS} +  
L_{q,3}^{W^+ - W^-, \rm NS} \Biggr] \nonumber\\
&&   
+ \Bigl[
  |V_{dc}|^2 (d - \overline{d}) 
+ |V_{sc}|^2 (s - \overline{s}) \Biggr] \otimes
H_{q,3}^{W^+ - W^-, \rm NS}~, 
\end{eqnarray}
%---------------------------------------------------------------------------------
with one massless Wilson coefficient $C_{q,3}^{W^+ - W^-, \rm NS}$ and two massive Wilson coefficients $L_{q,3}^{W^+ - 
W^-, \rm NS}$
and $H_{q,3}^{W^+ - W^-, \rm NS}$, see Section~\ref{sec:2}. The coefficients $V_{ij}$ are the Cabibbo-Kobayashi-Maskawa 
(CKM) \cite{Cabibbo:1963yz,Kobayashi:1973fv}
matrix elements, where $V_u = |V_{du}|^2 + |V_{su}|^2$, and the present numerical values are \cite{PDG}
%---------------------------------------------------------------------------------
\begin{eqnarray}
|V_{du}| &=& 0.97425  \\
|V_{su}| &=& 0.2253   \\ 
|V_{dc}| &=& 0.225    \\
|V_{sc}| &=& 0.986~.
\end{eqnarray}
%---------------------------------------------------------------------------------
The massless Wilson coefficient reads to $O(a_s^3)$
%---------------------------------------------------------------------------------
\begin{eqnarray}
C_{q,3}^{W^+ - W^-, \rm NS}\left(x,\frac{Q^2}{\mu^2}\right) = \delta(1-x) +\sum_{k=1}^3 a_s^k(\mu^2) C_{q,3}^{(k), W^+ - 
W^-, \rm 
NS}\left(x,\frac{Q^2}{\mu^2}\right)~, 
\label{eq:LQ3}
\end{eqnarray}
%---------------------------------------------------------------------------------
where $\mu$ denotes both the factorization and renormalization scale, which have been set equal
$\mu = \mu_R = \mu_F$\footnote{For the scale dependence of the Wilson coefficient see e.g. \cite{vanNeerven:2000uj}.}, and 
%---------------------------------------------------------------------------------
\begin{eqnarray}
A(x) \otimes B(x) = \int_0^1 dx_1 \int_0^1 dx_2 \delta(x - x_1 x_2) A(x_1) B(x_2)
\end{eqnarray}
%---------------------------------------------------------------------------------
is the Mellin convolution. $u, d$ and $s$ are the light quark number densities and $\bar{q}$ denotes the 
corresponding antiquark densities, written here for a proton target. The valence distributions are given by
%---------------------------------------------------------------------------------
\begin{eqnarray}
q_v = q - \bar{q}~.
\end{eqnarray}
%---------------------------------------------------------------------------------
Very often one considers the case 
%---------------------------------------------------------------------------------
\begin{eqnarray}
s_v = 0~.
\end{eqnarray}
%---------------------------------------------------------------------------------
Various experiments are carried out using isoscalar targets, possessing neutron-proton symmetry.
Here the following replacements have to be made
%---------------------------------------------------------------------------------
\begin{eqnarray}
u       &\rightarrow& \tfrac{1}{2} \left(u+d\right), \hspace*{3cm}
\bar{u} \rightarrow \tfrac{1}{2} \left(\bar{u}+\bar{d}\right)
\nonumber\\
d       &\rightarrow& \tfrac{1}{2} \left(u+d\right), \hspace*{3cm}
\bar{d} \rightarrow \tfrac{1}{2} \left(\bar{u}+\bar{d}\right).
\end{eqnarray}
%---------------------------------------------------------------------------------

The paper is organized as follows. In Section~\ref{sec:2} we calculate the heavy flavor contributions to the 
non-singlet Wilson coefficients in the asymptotic region $Q^2 \gg m^2$ to the structure function $xF_3^{W^+-W^-}(x,Q^2)$ 
to 3-loop order in the strong coupling constant. We present the results both in  Mellin $N$ and $x$--space.
Numerical results are given in Section~\ref{sec:3}. Consequences for the Gross-Llewellyn sum rule are discussed in
Section~\ref{sec:4}, and Section~\ref{sec:5} contains the conclusions. An appendix deals with a technical detail.
%%%%%%%%%%%%%%%%%%%%%%%%%%%%%%%%%%%%%%%%%%%%%%%%%%%%%%%%%%%%%%%%%%%%%%%%%%%%%%%%%%%%%%%%%%%%%%%%%%%
\section{The Wilson Coefficients}
\label{sec:2}
%%%%%%%%%%%%%%%%%%%%%%%%%%%%%%%%%%%%%%%%%%%%%%%%%%%%%%%%%%%%%%%%%%%%%%%%%%%%%%%%%%%%%%%%%%%%%%%%%%%

\vspace*{1mm}
\noindent
The asymptotic heavy flavor corrections to the structure function $xF_3^{W^+-W^-}(x,Q^2)$ are resulting from the  Wilson 
coefficients
$L_{q,3}^{W^+-W^-,\rm NS}$ and $H_{q,3}^{W^+-W^-,\rm NS}$. In the former case the charged current couples to a line of 
light 
quarks
only, while in the latter case a heavy quark is excited by the exchanged $W$-boson with or without other heavy flavor effects
through QCD corrections. In the following we work in Mellin space. The corresponding representation is obtained 
applying the transformation
%---------------------------------------------------------------------------------
\begin{eqnarray}
{\Mvec}[f(x)](N) = \int_0^1 dx x^{N-1} f(x)
\end{eqnarray}
%---------------------------------------------------------------------------------
to the above relations. 
One obtains
%---------------------------------------------------------------------------------
\begin{eqnarray}
L_{q,3}^{W^+ - W^-,\rm NS}(N_F+1) &=& 
A_{qq,Q}^{\rm NS} 
{C}_{q,3}^{W^+ - W^-, \rm NS}(N_F+1)
- {C}_{q,3}^{W^+ - W^-, \rm NS}(N_F)\\
H_{q,3}^{W^+ - W^-,\rm NS}(N_F+1) &=& 
A_{qq,Q}^{\rm NS} 
{C}_{q,3}^{W^+ - W^-, \rm NS}(N_F+1)
\end{eqnarray}
%---------------------------------------------------------------------------------
as the general relations. Here $A_{qq,Q}^{\rm NS} \equiv A_{qq,Q}^{\rm NS}(N_F+1)$ denotes the massive non-singlet
operator matrix element \cite{Bierenbaum:2009mv,Ablinger:2014vwa}.
The expansion of both the massive operator matrix element (OME) and the massless Wilson coefficients in the
strong coupling constant to 3-loop order leads to
%---------------------------------------------------------------------------------
\begin{eqnarray}
L_{q,3}^{W^+ - W^-,\rm NS}(N_F+1) &=& a_s^2 A_{qq,Q}^{(2),\rm NS} + \hat{C}_{q,3}^{(2),W^+ - W^-, \rm NS}(N_F) + 
a_s^3 \Biggl[A_{qq,Q}^{(3),\rm NS}  
\nonumber\\  &&
+ A_{qq,Q}^{(2),\rm NS} C_{q,3}^{(1),W^+ - W^-, \rm NS}(N_F+1) + \hat{C}_{q,3}^{(3),W^+ - W^-, \rm NS}(N_F)\Biggr]~, 
\label{eq:WIL1}
\\ 
H_{q,3}^{W^+ - W^-,\rm NS}(N_F+1) &=& 
1 + a_s C_{q,3}^{(1),W^+ - W^-, \rm NS}(N_F+1)
\nonumber\\ &&
+ a_s^2 \left[A_{qq,Q}^{(2),\rm NS} + C_{q,3}^{(2),W^+ - W^-, \rm NS}(N_F+1) \right]
\nonumber\\ &&
+ a_s^3 \Biggl[A_{qq,Q}^{(3),\rm NS} 
                + A_{qq,Q}^{(2),\rm NS} C_{q,3}^{(1),W^+ - W^-, \rm NS}(N_F+1) 
\nonumber\\ && 
+ {C}_{q,3}^{(3),W^+ - W^-, \rm NS}(N_F+1)\Biggr] 
\label{eq:WIL2}
\\
&=& 
L_{q,3}^{W^+ - W^-,\rm NS}(N_F+1) + {C}_{q,3}^{W^+ - W^-, \rm 
NS}(N_F),
\label{eq:H3}
\end{eqnarray}
%---------------------------------------------------------------------------------
with ${C}_{q,3}^{W^+ - W^-, \rm NS}(N_F)$ the massless Wilson coefficient up to 3-loop order.
Here we use the convention
%---------------------------------------------------------------------------------
\begin{eqnarray}
\hat{f}(N_F) =  f(N_F+1) - f(N_F)~.
\label{eq:HAT}
\end{eqnarray}
%---------------------------------------------------------------------------------
The notation  label `$N_F+1$' in some of the above quantities means that they depend on 
$N_F$ massless and one massive flavor.

The calculation of the different contributions to the Wilson coefficients (\ref{eq:WIL1}, 
\ref{eq:WIL2}) is performed 
in $D = 4 + \ep$ dimensions to regulate the Feynman integrals. The treatment of $\gamma_5$ has to be considered. In the flavor non-singlet case
both for the massive OMEs and the massless Wilson coefficients, $\gamma_5$ always appears in traces along one quark line.
For the asymptotic Wilson coefficient $L_{q,3}^{\rm NS}$, this line is massless, while for $H_{q,3}^{\rm NS}$ also 
heavy 
quarks
contribute due to single charged current flavor excitations like $s \rightarrow c$. Still we are considering the case in which power
corrections $(m^2/Q^2)^k,~k \in \mathbb{N}, k \geq 1,$ are disregarded and therefore the corresponding 
line has to be treated as 
massless. In this case 
a (reversed) Ward-Takahashi identity implies anti-commuting $\gamma_5$, mapping the corresponding vertex correction to a 
self-energy correction, which is the same in the case without $\gamma_5$. 

The inclusive massive OME  $A_{qq,Q}^{\rm NS}$ to 3-loop order for even and odd moments $N$ has been calculated
in Ref.~\cite{Ablinger:2014vwa}. The corresponding Feynman integrals have been reduced using integration-by-parts 
relations
\cite{IBP} applying an extension of the package {\tt Reduze\;2} \cite{REDUZE2}\footnote{The package {\tt Reduze\;2} 
uses the packages {\tt Fermat} \cite{FERMAT} and {\tt Ginac} \cite{Bauer:2000cp}.}. The master integrals have been calculated
using hypergeometric, Mellin-Barnes and differential equation techniques, mapping them to recurrences, which have been solved
by modern summation technologies using extensively the packages {\tt Sigma} \cite{SIG1,SIG2}, {\tt EvaluateMultiSums}, 
{\tt SumProduction} \cite{EMSSP}, {\tt $\rho$sum} \cite{RHOSUM}, and {\tt HarmonicSums} \cite{HARMONICSUMS}.

In Mellin $N$ space the Wilson coefficient can be expressed by nested harmonic sums
$S_{\vec{a}}(N)$ \cite{HSUM} which are defined by
%---------------------------------------------------------------------------------------------------------------
\begin{eqnarray}
S_{b,\vec{a}}(N) = \sum_{k=1}^N \frac{({\rm sign}(b))^k}{k^{|b|}} S_{\vec{a}}(k),~~~S_\emptyset = 1,~b, a_i 
\in
\mathbb{Z}, b, a_i \neq 0, N > 0, N \in \mathbb{N}.
\end{eqnarray}
%------------------------------------------------------------------------------------------------------------------------------
In the following, we drop the argument $N$ of the harmonic sums and use the short-hand
notation $S_{\vec{a}}(N) \equiv S_{\vec{a}}$. The Wilson coefficients depend on the logarithms
%---------------------------------------------------------------------------------------------------------------
\begin{eqnarray}
L_Q = \ln\left(\frac{Q^2}{\mu^2}\right)~~~~~~~\text{and}~~~~~~L_M = \ln\left(\frac{m^2}{\mu^2}\right).
\end{eqnarray}
%------------------------------------------------------------------------------------------------------------------------------

As a short-hand notation we define the leading order splitting function $\gamma_{qq}^{(0)}$  without its color factor 
%-----------------------------------------------------------------------------------------------------------------------
\begin{eqnarray}
\gamma_{qq}^{(0)} = 4 \left[2 S_1 - \frac{3 N^2+3 N+2}{2 N (N+1)}\right]~.
\end{eqnarray}
%-----------------------------------------------------------------------------------------------------------------------
The massive Wilson coefficient for the structure function $xF_3^{W^+ - W^-}(x,Q^2)$ in the asymptotic region in 
the on-shell scheme in Mellin space is given by 
%------------------------------------------------------------------------------------------------------------------------------

%-----------------------------------------------------------------------------------------------------
where $\hat{C}_{q,3}^{\mathrm{NS},(3)}(N_F)$ denotes the contribution due to the massless Wilson coefficient at 3-loop 
order \cite{Moch:2008fj}, and the polynomials $P_i$ are
%-----------------------------------------------------------------------------------------------------
\begin{eqnarray}
P_1    &=& -17 N^4-34 N^3-29 N^2-12 N-24 \\
P_2    &=& -3 N^4-6 N^3-47 N^2-20 N+12 \\
P_3    &=& N^4+2 N^3-N^2-2 N-4 \\
P_4    &=& 3 N^4+6 N^3+47 N^2+20 N-12 \\
P_5    &=& 7 N^4+14 N^3+3 N^2-4 N-4 \\
P_6    &=& 19 N^4+38 N^3-9 N^2-20 N+4 \\
P_7    &=& 28 N^4+56 N^3+28 N^2+2 N+1 \\
P_8    &=& 33 N^4+54 N^3+9 N^2-52 N-28 \\
P_9    &=& 57 N^4+96 N^3+65 N^2-10 N-24 \\
P_{10} &=& 112 N^4+224 N^3+121 N^2+9 N+9 \\
P_{11} &=& 141 N^4+246 N^3+241 N^2-8 N-84 \\
P_{12} &=& 181 N^4+266 N^3+82 N^2-3 N+18 \\
P_{13} &=& 235 N^4+524 N^3+211 N^2+30 N+72 \\
P_{14} &=& 359 N^4+772 N^3+335 N^2+30 N+72 \\
P_{15} &=& 501 N^4+894 N^3+541 N^2-116 N-204 \\
P_{16} &=& 561 N^4+1122 N^3+767 N^2+302 N+48 \\
P_{17} &=& 1131 N^4+2118 N^3+1307 N^2+32 N-276 \\
P_{18} &=& 1139 N^4+2710 N^3+635 N^2+216 N+828 \\
P_{19} &=& 1199 N^4+2398 N^3+1181 N^2+18 N+90 \\
P_{20} &=& 1220 N^4+2359 N^3+1934 N^2+357 N-138 \\
P_{21} &=& 3 N^5+11 N^4+10 N^3+19 N^2+23 N+16 \\
P_{22} &=& 12 N^5+16 N^4+18 N^3-15 N^2-5 N-8 \\
P_{23} &=& 27 N^5+863 N^4+1573 N^3+1151 N^2+144 N-36 \\
P_{24} &=& 648 N^5-2103 N^4-4278 N^3-3505 N^2-682 N-432 \\
P_{25} &=& -11145 N^6-32355 N^5-37523 N^4-14329 N^3+1240 N^2-1032 N-2088 \\
P_{26} &=& -151 N^6-469 N^5-181 N^4+305 N^3+80 N^2-88 N-56 \\
P_{27} &=& N^6+3 N^5-8 N^4-21 N^3-23 N^2-12 N-4 \\
P_{28} &=& 6 N^6+18 N^5-N^4-20 N^3+46 N^2+29 N-6 \\
P_{29} &=& 15 N^6+36 N^5+30 N^4-24 N^3+3 N^2+16 N+20 \\
P_{30} &=& 155 N^6+465 N^5+465 N^4+155 N^3+108 N^2+108 N+54 \\
P_{31} &=& 216 N^6+567 N^5+687 N^4+285 N^3+37 N^2-44 N+12 \\
P_{32} &=& 309 N^6+807 N^5+693 N^4-463 N^3-638 N^2+68 N+216 \\
P_{33} &=& 525 N^6+1575 N^5+1535 N^4+973 N^3+536 N^2+48 N-72 \\
P_{34} &=& 609 N^6+1485 N^5+1393 N^4+83 N^3-422 N^2+156 N+216 \\
P_{35} &=& 795 N^6+2043 N^5+2075 N^4+517 N^3-298 N^2+156 N+216 \\
P_{36} &=& 868 N^6+2469 N^5+2487 N^4+940 N^3+171 N^2+207 N+144 \\
P_{37} &=& 1407 N^6+3825 N^5+4211 N^4+1783 N^3-250 N^2-240 N+144 \\
P_{38} &=& 1770 N^6+4671 N^5+4765 N^4+1205 N^3-227 N^2+1044 N+756 \\
P_{39} &=& 7531 N^6+23673 N^5+23055 N^4+7375 N^3+1614 N^2+936 N-324 \\
P_{40} &=& -4785 N^8-19140 N^7-18754 N^6+1320 N^5+12723 N^4+6548 N^3+4080 N^2
\nonumber\\ &&
-648 N-1728 \\
P_{41} &=& -45 N^8-162 N^7-858 N^6-936 N^5-1629 N^4-1094 N^3-804 N^2
\nonumber\\ &&
-40 N+192 \\
P_{42} &=& N^8+4 N^7+13 N^6+25 N^5+57 N^4+77 N^3+55 N^2+20 N+4 \\
P_{43} &=& 3 N^8+12 N^7+16 N^6+6 N^5+30 N^4+64 N^3+73 N^2+40 N+12 \\
P_{44} &=& 3549 N^8+14196 N^7+23870 N^6+25380 N^5+15165 N^4+1712 N^3-2016 N^2
\nonumber\\ &&
+144 N+432 \\
P_{45} &=& 5487 N^8+21948 N^7+36370 N^6+28836 N^5+11943 N^4+4312 N^3+2016 N^2
\nonumber\\ &&
-144 N-432 \\
P_{46} &=& 10807 N^8+43228 N^7+63222 N^6+40150 N^5+14587 N^4+9018 N^3+7452 N^2
\nonumber\\ &&
+2376 N+324 \\
P_{47} &=& 42591 N^8+166764 N^7+245664 N^6+129982 N^5-13295 N^4-25978 N^3
\nonumber\\ &&
+3560 N^2-3192 N-4464 \\
P_{48} &=& -18351 N^{10}-89784 N^9-210021 N^8-271638 N^7-219369 N^6-90572 N^5
\nonumber\\ &&
-26491 N^4-7790 N^3-1992 N^2-2760 N-2160 \\
P_{49} &=& 165 N^{10}+825 N^9+109664 N^8+331682 N^7+457641 N^6+346145 N^5
\nonumber\\ &&
+219290 N^4+86724 N^3+13608 N^2+14256 N+10368 \\
P_{50} &=& 828 N^{10}+3492 N^9+4305 N^8-2013 N^7-8540 N^6-3822 N^5-1157 N^4
\nonumber\\ &&
-3057 N^3-4112 N^2-324 N+576 \\
P_{51} &=& 8274 N^{10}+37149 N^9+53630 N^8+7538 N^7-59902 N^6-55159 N^5
\nonumber\\ &&
-6994 N^4+3272 N^3-9048 N^2-1656 N+2160~.
\end{eqnarray}

%------------------------------------------------------------------------------------------------------------------------------
Here $C_A = N_c, C_F = (N_c^2-1)/(2 N_c), T_F =1/2$ and $d^{abc} d^{abc}/N_c = (N_c^2-1)(N_c^2-4)/N_c^2$ denote the 
color factors 
for $SU(N_c)$, with $N_c = 3$ in case of QCD. The constant $B_4$ is given by
%---------------------------------------------------------------------------------------------------------------
\begin{eqnarray}
B_4 = - 4 \zeta_2 \ln^2(2) + \frac{2}{3} \ln^4(2) - \frac{13}{2} \zeta_4 + 16 \Li_4\left(\frac{1}{2}\right),
\end{eqnarray}
%------------------------------------------------------------------------------------------------------------------------------
which is related to multiple zeta values \cite{Blumlein:2009cf}, $\zeta_n = \sum_{k=1}^\infty 1/k^n,~~n \in \mathbb{N}, n 
\geq 2$, denote the values of Riemann's $\zeta$-function at integer argument and $\Li_n(x), n \in \mathbb{N}$, is the 
polylogarithm \cite{POLY}. Terms containing the color factor $d^{abc} d^{abc}/N_c$ stem only from the massless Wilson 
coefficient and the non-singlet valence anomalous dimension at 3-loop order \cite{NSANOM}, cf.~Appendix~A.

One obtains the analytic continuation of the harmonic sums to complex values of $N$ by performing their asymptotic expansion
analytically, cf.~\cite{Blumlein:2009ta,Blumlein:2009fz}. These expansions can be obtained automatically using the 
package {\tt HarmonicSums} \cite{HARMONICSUMS}. Furthermore, the nested harmonic sums obey the shift relations 
%---------------------------------------------------------------------------------------------------------------
\begin{eqnarray}
S_{b,\vec{a}}(N) = S_{b,\vec{a}}(N-1) + \frac{({\rm sign}(b))^N}{N^{|b|}} S_{\vec{a}}(N)~, 
\end{eqnarray}
%------------------------------------------------------------------------------------------------------------------------------
through which any regular point in the complex plane can be reached using the analytic asymptotic representation as input.
The poles of the nested harmonic sums $S_{\vec{a}}(N)$ are located at the non-positive integers. In data analyses,
one may thus encode the QCD evolution \cite{Blumlein:1997em} together with the Wilson coefficient for complex values of $N$
analytically and finally perform one numerical contour integral around the singularities of the problem.\footnote{For precise
numerical implementations of the analytic continuation of harmonic sums see \cite{ANCONT}.}

In $x$-space the Wilson coefficient is represented in terms of harmonic polylogarithms \cite{Remiddi:1999ew} over the alphabet
$\{f_0, f_1, f_{-1}\}$, which were again reduced applying the shuffle relations \cite{Blumlein:2003gb}. They are defined by
%---------------------------------------------------------------------------------------------------------------
\begin{eqnarray}
H_{b,\vec{a}}(x) &=& \int_0^x dy f_b(y) H_{\vec{a}}(y),~~~H_{\scriptsize \underbrace{0,...,0}_{k}}(x) = \frac{1}{k!} 
\ln^k(x),~~~H_\emptyset 
= 1,\\
f_0(x) &=& \frac{1}{x},~~~~~~f_1(x) = \frac{1}{1-x},~~~~~~f_{-1}(x) = \frac{1}{1+x}~.
\end{eqnarray}
%------------------------------------------------------------------------------------------------------------------------------
The Wilson coefficient is represented by three contributions, the $(...)_+$-distribution, the $\delta(1-x)$-term,
and the regular term, with the $+$-distribution being defined by    
%---------------------------------------------------------------------------------------------------------------
\begin{eqnarray}
\int_0^1 dy \left[F(y)\right]_+ g(y) = \int_0^1 dy F(y) \left[g(y) - g(1)\right]~.
\end{eqnarray}
%------------------------------------------------------------------------------------------------------------------------------
\noindent
It reads

%------------------------------------------------------------------------------------------------------------------------------
\newpage


%------------------------------------------------------------------------------------------------------------------------------
Again, we used the short hand notation $H_{\vec{a}}(x) \equiv H_{\vec{a}}$ also here.
The transformation of the Wilson coefficient to the $\overline{\rm MS}$ scheme for the 
heavy quark mass affects the massive 
OME at 3-loops and was given in Ref.~\cite{Ablinger:2014vwa}; the terms are the same in the 
unpolarized and polarized case.

The $+$-distribution of the Wilson coefficient also contains $1/(1-x)^2$-terms, cf.~\cite{Ablinger:2014vwa}. 
The explicit expressions
for the Wilson coefficient $H_{q,3}^{\rm NS}(N_F+1)$ can be easily obtained from Eq.~(\ref{eq:H3}).

%%%%%%%%%%%%%%%%%%%%%%%%%%%%%%%%%%%%%%%%%%%%%%%%%%%%%%%%%%%%%%%%%%%%%%%%%%%%%%%%%%%%%%%%%%%%%%%%%%%
\section{Numerical Results}
\label{sec:3}
%%%%%%%%%%%%%%%%%%%%%%%%%%%%%%%%%%%%%%%%%%%%%%%%%%%%%%%%%%%%%%%%%%%%%%%%%%%%%%%%%%%%%%%%%%%%%%%%%%%

\vspace*{1mm}
\noindent
In what follows, we will choose the factorization and renormalization scale $\mu^2 = Q^2$. We first study the 
behaviour of the massive and massless Wilson coefficients in the small and large $x$ region and then give numerical 
illustrations in the whole $x$-region.
 
At small $x$, the pure massive Wilson coefficient behaves like
%-----------------------------------------------------------------------------------------------------------------------------
\begin{eqnarray}
L_{q,3}^{\rm NS}(N_F+1) - \hat{C}_{q,3}^{\mathrm{NS}}(N_F)
&\propto& a_s^2 \frac{2}{3} C_F T_F \ln^2(x) + a_s^3 \Biggl[\frac{16}{27} C_A C_F T_F  - \frac{5}{9} C_F^2 T_F \Biggr] 
\ln^4(x), 
\nonumber\\
\end{eqnarray}
%-----------------------------------------------------------------------------------------------------------------------------
while in the region $x \rightarrow 1$ one obtains\footnote{There is a typo in the second contribution to the $O(a_s^3)$ term
of Eq. (3.2) in \cite{Behring:2015zaa}, which should be $\propto \ln(Q^2/m^2)$.}
%-----------------------------------------------------------------------------------------------------------------------------
\begin{eqnarray}
L_{q,3}^{\rm NS}(N_F+1) -\hat{C}_{q,3}^{\mathrm{NS}}(N_F)
&\propto& 
a_s^2 C_F T_F \left[\frac{224}{27}  + \frac{80}{9} L_M + \frac{8}{3} L_M^2\right] \left(\frac{1}{1-x}\right)_+
\nonumber\\ &&
+ a_s^3 C_F^2 T_F \Biggl[\frac{448}{9}  
+ \frac{160}{3} L_M
+ 16 L_M^2 \Biggr] \left(\frac{\ln^2(1-x)}{1-x}\right)_+~.
\nonumber\\ 
\end{eqnarray}
%-----------------------------------------------------------------------------------------------------------------------------
Likewise one obtains for $H_{q,3}^{\rm NS}(N_F+1)$ at small $x$ 
%-----------------------------------------------------------------------------------------------------------------------------
\begin{eqnarray}
H_{q,3}^{\rm NS}(N_F+1) - {C}_{q,3}^{\mathrm{NS}}(N_F+1)
&\propto& a_s^2 \frac{2}{3} C_F T_F \ln^2(x) + a_s^3 \Biggl[\frac{16}{27} C_A C_F T_F  - \frac{5}{9} C_F^2 T_F \Biggr] 
\ln^4(x),
\nonumber\\ 
\end{eqnarray}
%-----------------------------------------------------------------------------------------------------------------------------
and in the region $x \rightarrow 1$
%-----------------------------------------------------------------------------------------------------------------------------
\begin{eqnarray}
H_{q,3}^{\rm NS}(N_F+1) - {C}_{q,3}^{\mathrm{NS}}(N_F+1)
&\propto& 
a_s^2 C_F T_F \left[\frac{224}{27}  + \frac{80}{9} L_M + \frac{8}{3} L_M^2\right] \left(\frac{1}{1-x}\right)_+
\nonumber\\ &&
+ a_s^3 C_F^2 T_F \Biggl[\frac{448}{9}
+ \frac{160}{3} L_M
+ 16 L_M^2 \Biggr] \left(\frac{\ln^2(1-x)}{1-x}\right)_+~.
\nonumber\\
\end{eqnarray}
%-----------------------------------------------------------------------------------------------------------------------------

The above results can be compared with the case of the massless Wilson coefficient $\hat{C}_{q,3}$ and ${C}_{q,3}$, 
respectively. The following dominant terms are found in the small and large $x$ regions,
%-----------------------------------------------------------------------------------------------------------------------------
\begin{eqnarray}
\hat{C}_{q,3}^{\mathrm{NS},(2)}(N_F) &\propto& a_s^2 \frac{10}{3} C_F T_F \ln^2(x)
\\
\hat{C}_{q,3}^{\mathrm{NS},(3)}(N_F)
&\propto& - a_s^3 \frac{2}{15} \frac{d^{abc} d^{abc}}{N_c}  \ln^5(x)
\\                               
{C}_{q,3}^{\mathrm{NS},(1)}(N_F) &\propto& - a_s 2 C_F  \ln(x)
\\
{C}_{q,3}^{\mathrm{NS},(2)}(N_F) &\propto&  a_s^2 \left[\frac{7}{3} C_F^2 - 2 C_A C_F \right] \ln^3(x)
\\
{C}_{q,3}^{\mathrm{NS},(3)}(N_F) &\propto&  a_s^3 \left[\frac{2}{5} C_A^2 C_F - \frac{29}{15} C_A C_F^2 
+ \frac{53}{30} C_F^3 - \frac{2}{15} \frac{d^{abc} d^{abc}}{N_c} N_F \right] \ln^5(x) 
\end{eqnarray}
%-----------------------------------------------------------------------------------------------------------------------------
and
%-----------------------------------------------------------------------------------------------------------------------------
\begin{eqnarray}
\hat{C}_{q,3}^{\mathrm{NS},(2)}(N_F) &\propto& a_s^2 \frac{8}{3} C_F T_F \left(\frac{\ln^2(1-x)}{1-x}\right)_+
\\
\hat{C}_{q,3}^{\mathrm{NS},(3)}(N_F)
&\propto& a_s^3 \frac{80}{9} C_F^2 T_F \left(\frac{\ln^4(1-x)}{1-x}\right)_+
\\                               
{C}_{q,3}^{\mathrm{NS},(1)}(N_F) &\propto& a_s 4 C_F  \left(\frac{\ln(1-x)}{1-x}\right)_+
\\
{C}_{q,3}^{\mathrm{NS},(2)}(N_F) &\propto&  a_s^2 8 C_F^2  \left(\frac{\ln^3(1-x)}{1-x}\right)_+
\\
{C}_{q,3}^{\mathrm{NS},(3)}(N_F) &\propto&  a_s^3 8 C_F^3 
\left(\frac{\ln^5(1-x)}{1-x}\right)_+~, 
\end{eqnarray}
%-----------------------------------------------------------------------------------------------------------------------------
respectively.

Let us finally also consider the case $L_Q \neq 0$. Terms of this order have the following leading small and large $x$ 
behaviour
%-----------------------------------------------------------------------------------------------------------------------------
\begin{eqnarray}
\hat{C}_{q,3}^{\mathrm{NS},(2)}(N_F) &\propto& - a_s^2 \frac{16}{3} C_F T_F L_Q \ln(x)
\\
\hat{C}_{q,3}^{\mathrm{NS},(3)}(N_F)
&\propto&  a_s^3 \frac{1}{3} \frac{d^{abc} d^{abc}}{N_c} L_Q \ln^4(x)
\\                               
{C}_{q,3}^{\mathrm{NS},(1)}(N_F) &\propto&  a_s 2 C_F L_Q 
\\
{C}_{q,3}^{\mathrm{NS},(2)}(N_F) &\propto&  a_s^2 4 C_F \left[C_A - C_F\right] L_Q \ln^2(x)
\\
{C}_{q,3}^{\mathrm{NS},(3)}(N_F) &\propto&  a_s^3 \left[-C_A^2 C_F + \frac{17}{3} C_A C_F^2 
- \frac{11}{2} C_F^3 + \frac{1}{3} \frac{d^{abc} d^{abc}}{N_c} N_F \right] L_Q \ln^4(x) 
\end{eqnarray}
%-----------------------------------------------------------------------------------------------------------------------------
and
%-----------------------------------------------------------------------------------------------------------------------------
\begin{eqnarray}
\hat{C}_{q,3}^{\mathrm{NS},(2)}(N_F) &\propto& a_s^2 \frac{16}{3} C_F T_F L_Q \left(\frac{\ln(1-x)}{1-x}\right)_+
\\
\hat{C}_{q,3}^{\mathrm{NS},(3)}(N_F)
&\propto& a_s^3 \frac{320}{9} C_F^2 T_F L_Q \left(\frac{\ln^3(1-x)}{1-x}\right)_+
\\                               
{C}_{q,3}^{\mathrm{NS},(1)}(N_F) &\propto& a_s 4 C_F  \left(\frac{1}{1-x}\right)_+
\\
{C}_{q,3}^{\mathrm{NS},(2)}(N_F) &\propto&  a_s^2 24 C_F^2  \left(\frac{\ln^2(1-x)}{1-x}\right)_+
\\
{C}_{q,3}^{\mathrm{NS},(3)}(N_F) &\propto&  a_s^3 40 C_F^3 L_Q 
\left(\frac{\ln^4(1-x)}{1-x}\right)_+,
\end{eqnarray}
%-----------------------------------------------------------------------------------------------------------------------------
respectively, and are less singular than those for $L_Q = 0$.
%---------------------------------------------------------------------------------------------------------------------------------
\begin{figure}[H]
\centering
\includegraphics[width=0.8\textwidth]{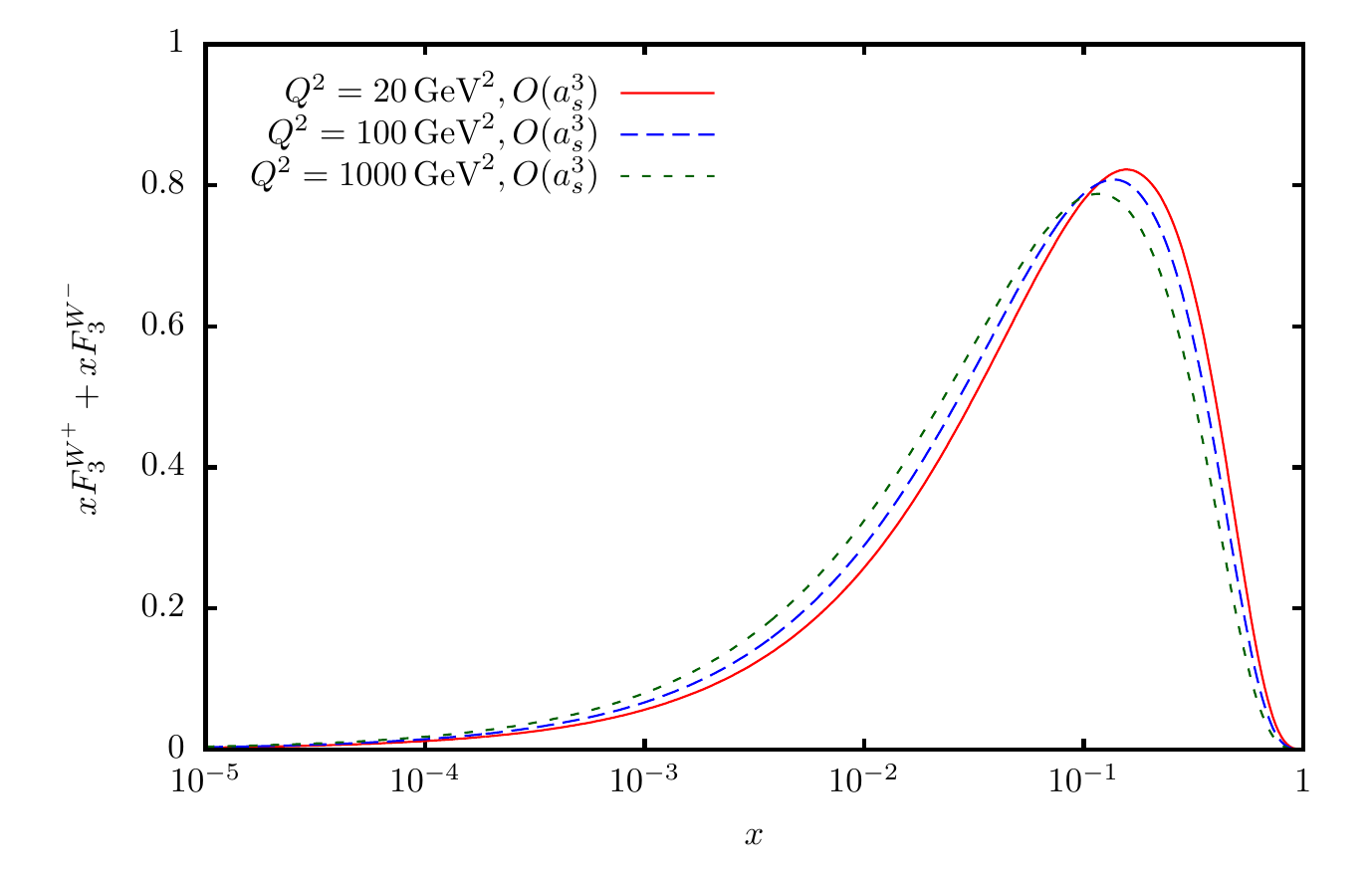}  
\caption{\sf \small The corrections up to 3-loop order to the combination of the structure functions
$xF_3^{W^+}(x,Q^2) + xF_3^{W^-}(x,Q^2)$ off a proton target, including both the massless
and the charm contributions in the asymptotic approximation in the on-shell scheme for $m_c = 
1.59~\GeV$ \cite{Alekhin:2014sya} as a function of $x$ and $Q^2$. The parton distribution 
functions of Ref.~\cite{Alekhin:2013nda} have been used with $\alpha_s(M_Z^2) = 0.1132$. 
These settings are the same in the subsequent Figures.}
\label{fig:1} 
\end{figure}
%---------------------------------------------------------------------------------------------------------------------------------

The small $x$ behaviour can be compared with leading order predictions for the non-singlet 
%---------------------------------------------------------------------------------------------------------------------------------
\begin{figure}[H]
\centering
\includegraphics[width=0.8\textwidth]{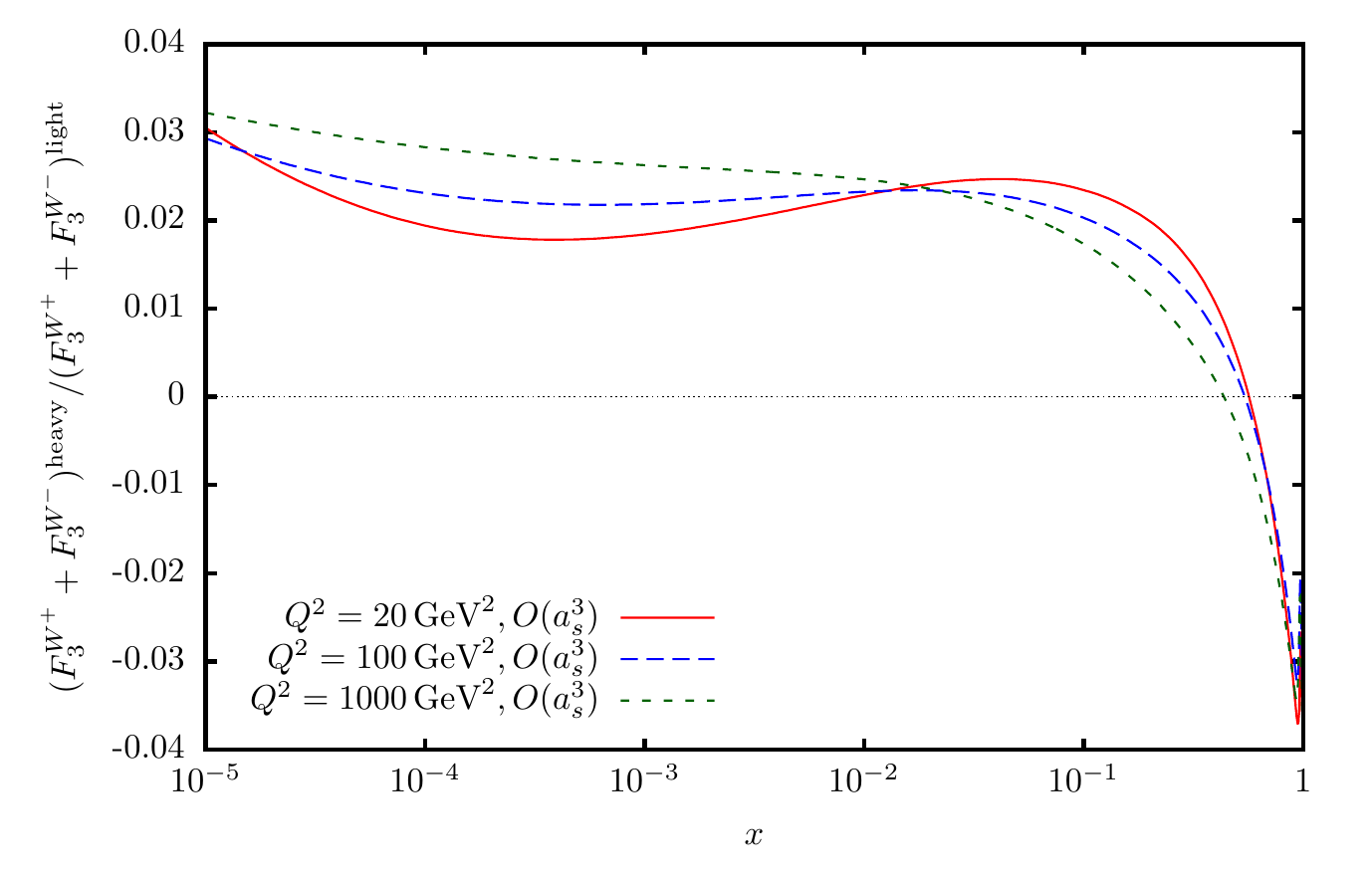}  
\caption{\sf \small Ratio of the structure functions $xF_3^{W^+}(x,Q^2) + 
xF_3^{W^-}(x,Q^2)$ off a proton target up to 3-loop order for the charm contribution and the massless 
terms for three 
flavors.} 
\label{fig:2} 
\end{figure}
%---------------------------------------------------------------------------------------------------------------------------------
%---------------------------------------------------------------------------------------------------------------------------------
\begin{figure}[H]
\centering
\includegraphics[width=0.8\textwidth]{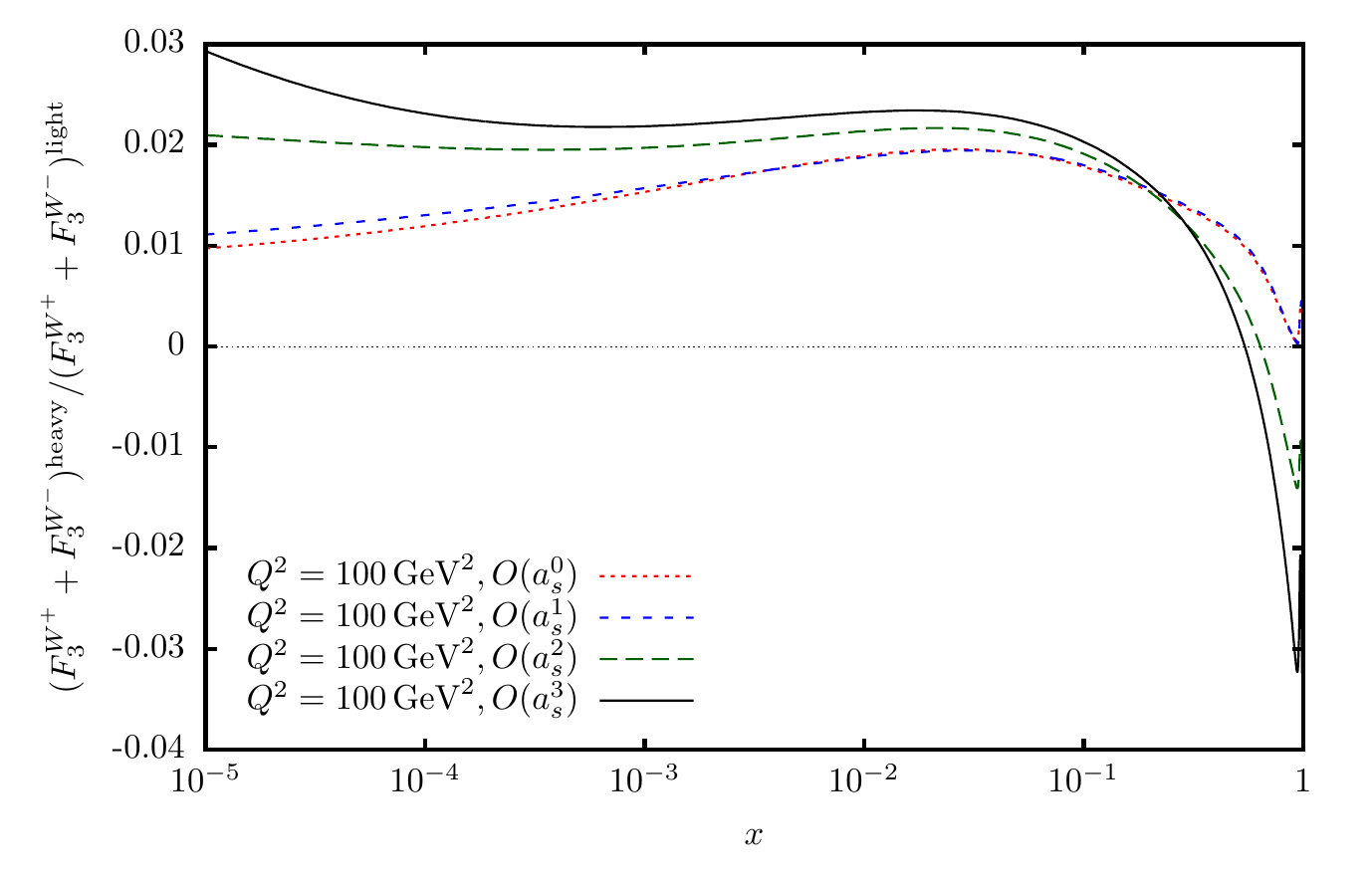}  
\caption{\sf \small Ratio of the structure functions $xF_3^{W^+}(x,Q^2) + xF_3^{W^-}(x,Q^2)$ off a proton target up to 
3-loop order for the charm contribution and the massless terms for three 
flavors from tree-level to $O(a_s^3)$ at $Q^2 = 100~\GeV^2$.}
\label{fig:3} 
\end{figure}
%---------------------------------------------------------------------------------------------------------------------------------

\noindent
evolution kernel 
in Refs.~\cite{Kirschner:1983di,Blumlein:1995jp}. Indeed, both the massive and massless contributions follow 
the principle pattern $\sim c_k a_s^{k+1} \ln^{2k}(x)$. However, as is well known \cite{Blumlein:1995jp}, less 
singular terms widely cancel the numerical effect of these leading terms. For the large $x$ terms 
the massless terms exhibit a stronger soft singularity than the massive ones. 

%---------------------------------------------------------------------------------------------------------------------------------
\begin{figure}[H]
\centering
\includegraphics[width=0.8\textwidth]{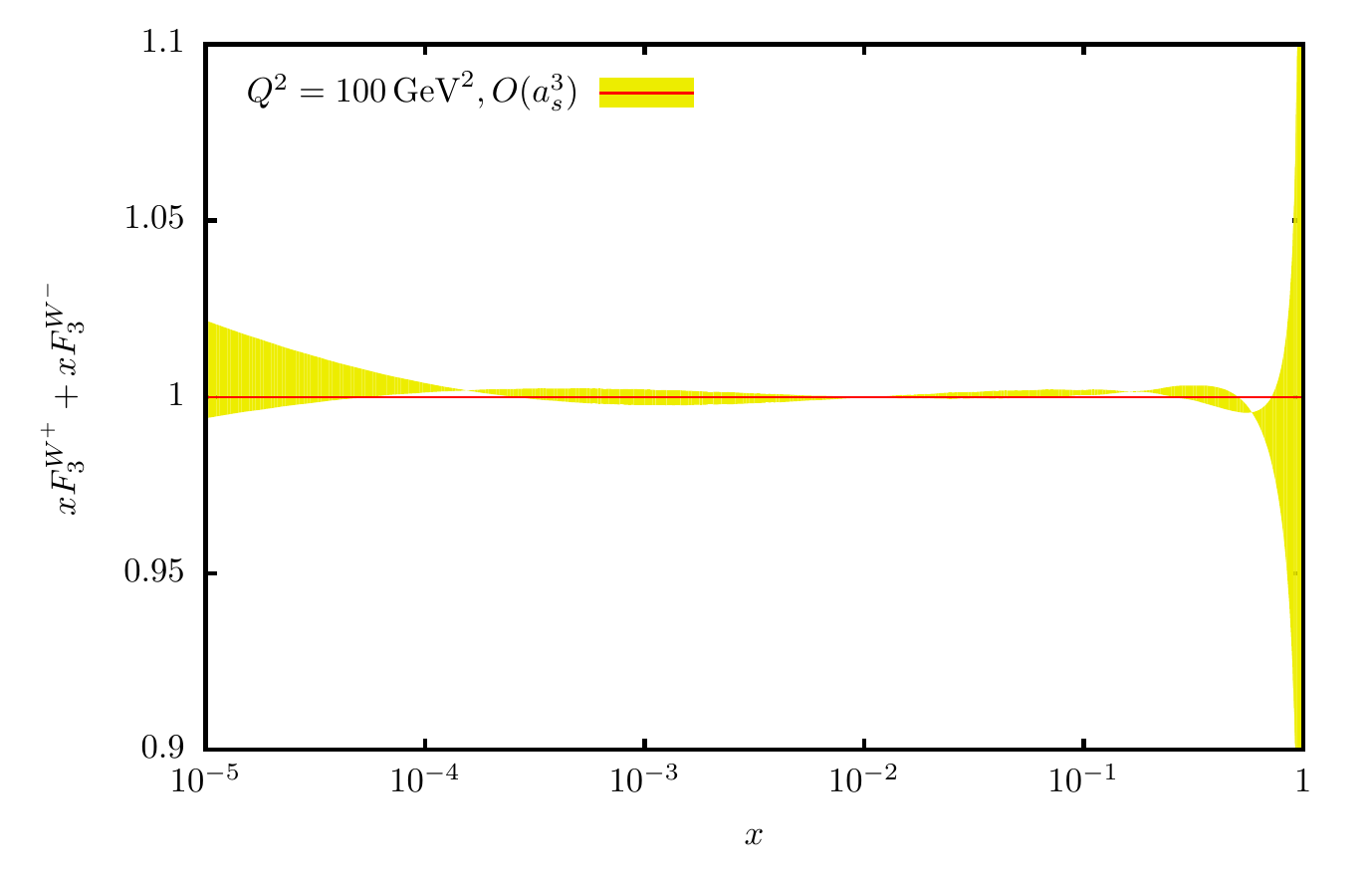}  
\caption{\sf \small The ratio of the structure functions $xF_3^{W^+}(x,\mu^2) +
xF_3^{W^-}(x,\mu^2)$ to $xF_3^{W^+}(x,Q^2) +
xF_3^{W^-}(x,Q^2)$, varying $\mu^2 \in [Q^2/4, 4 Q^2]$ at $O(a_s^3)$ and $Q^2 = 100~\GeV^2$ as a function of $x$.}
\label{fig:4} 
\end{figure}
%---------------------------------------------------------------------------------------------------------------------------------
%---------------------------------------------------------------------------------------------------------------------------------
\begin{figure}[H]
\centering
\includegraphics[width=0.8\textwidth]{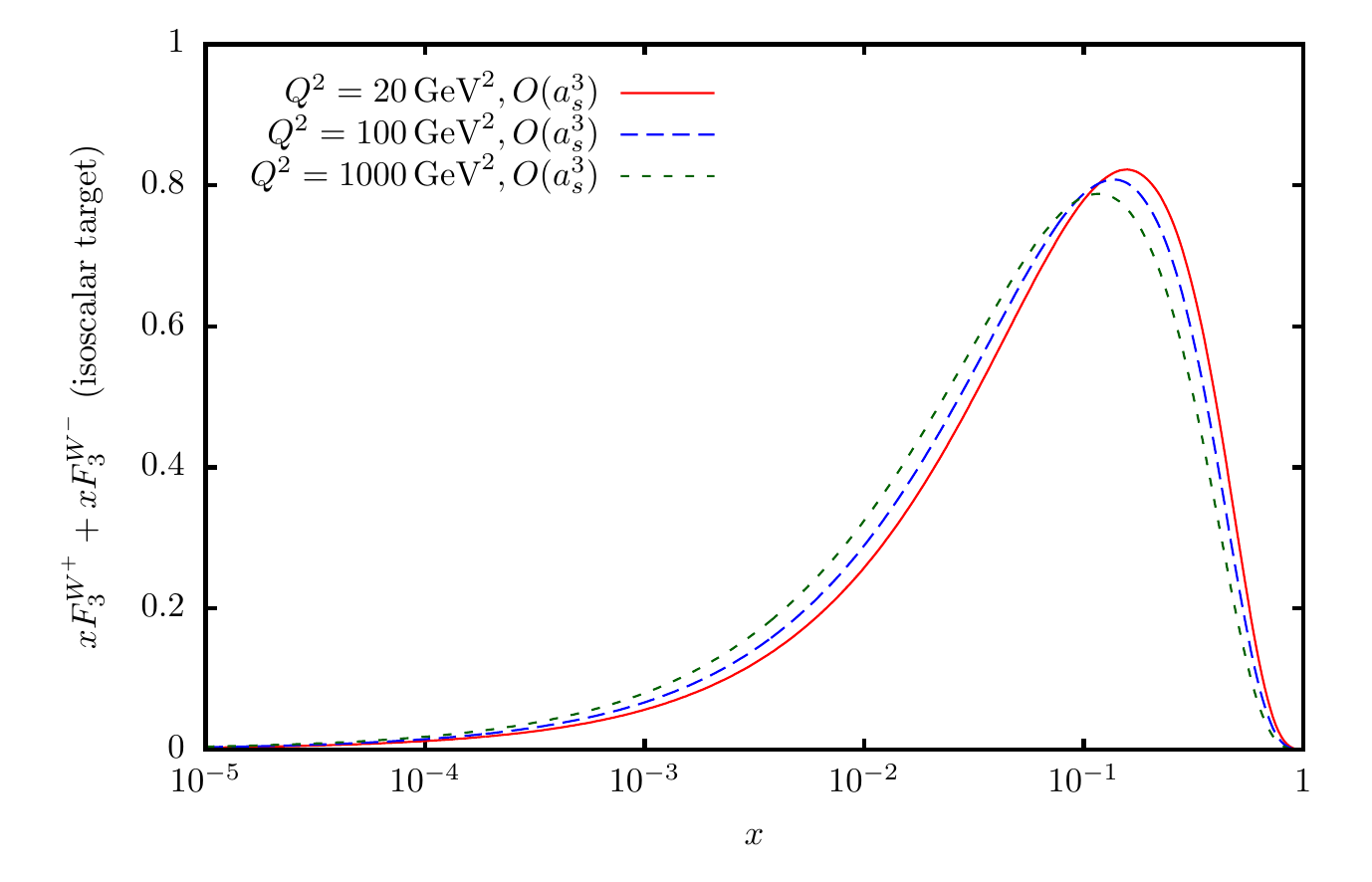}  
\caption{\sf \small The corrections up to 3-loop order to the combination of the structure functions
$xF_3^{W^+}(x,Q^2) + xF_3^{W^-}(x,Q^2)$ off a nucleon in an isoscalar target, including both the massless
and charm contributions in the asymptotic approximation.}
\label{fig:5} 
\end{figure}
%---------------------------------------------------------------------------------------------------------------------------------

\noindent
In Figure~1 we present the combination of structure functions $xF_3^{W^+ - W^-, \rm NS}$ up to $O(a_s^3)$ at a proton 
target 
as a function of $x$ and $Q^2$ up to 3-loop order for three massless flavors and charm in the asymptotic 
representation, assuming $m_c = 1.59~\GeV$ \cite{Alekhin:2012vu} in the on-shell scheme for mass 
\noindent
renormalization and 
setting $\mu^2 = Q^2$. It shows valence-like scaling violations moving towards smaller values of $x$.

The effect of the 
charm contributions are illustrated in comparison to the purely massless ones up to $O(a_s^3)$ in Figure~2 evolving with 
$Q^2$. 
The contribution amounts to a correction of up to $+3\%$ in the small $x$ region and to $-3\%$ in the large $x$ region 
with some evolution in $Q^2$. 

In Figure~3 we show a similar ratio as in Figure~2, but taken for each order of the perturbative corrections to 
$xF_3^{W^+}(x,\mu^2) + xF_3^{W^-}(x,\mu^2)$  separately at $Q^2 = 100~\GeV^2$. In the small $x$ region the corrections 
grow from 1\% to 3\%. At large 
values of $x$ the relative corrections tend to become negative and do also amount to a few per cent.
Figure~4 shows the remaining scale variations of the combination $xF_3^{W^+}(x,\mu^2) + xF_3^{W^-}(x,\mu^2)$ up to 3-loop 
order in the region $\mu^2 \in [Q^2/4, 4 Q^2]$ normalized to the value at $\mu^2 = Q^2$ (yellow band) as a function of $x$ 
for $Q^2 = 100~\GeV^2$. The behaviour is overall flat with $\pm 1\%$ variations in the medium $x$-range and shows larger 
uncertainties at very low and large values of $x$. The behaviour is similar at other virtualities.

Figure~5 shows the combination of structure functions $xF_3^{W^+ - W^-, \rm NS}$ up to $O(a_s^3)$ at a nucleon in an 
isoscalar target.
The differences to the case of the proton target shown in Figure~1 turn out to be rather small in the case of this 
combination, which widely pronounces isoscalarity due to the present combination of currents, if $s_v \approx  0$.
%-----------------------------------------------------------------------------------------------------------------------
\section{The Gross-Llewellyn Smith Sum Rule}
\label{sec:4}
%-----------------------------------------------------------------------------------------------------------------------

\vspace*{1mm}
\noindent
The Gross-Llewellyn Smith sum rule \cite{Gross:1969jf} refers to the first moment of the flavor non-singlet combination
%---------------------------------------------------------------------------
\begin{eqnarray}
\label{eq:PBJ}
\int_0^1 {dx} \left[F_3^{\bar{\nu} p}(x,Q^2) + F_3^{{\nu} p}(x,Q^2)\right] 
= 6 C_{\rm GLS}(\hat{a}_s),
\end{eqnarray}
%-----------------------------------------------------------------------------------------------------------------------------
with $\hat{a}_s = \alpha_s/\pi$ and idealized CKM mixing. The 1-loop 
\cite{Bardeen:1978yd,Altarelli:1978id,Humpert:1980uv,Furmanski:1981cw}, 2-loop 
\cite{Gorishnii:1985xm}, 3-loop \cite{Larin:1991tj} and 4-loop QCD corrections \cite{Baikov:2012zn,Baikov:2010je} in the 
massless case are given by
%-----------------------------------------------------------------------------------------------------------------------------
\begin{eqnarray}
\label{eq:BJSR}
C_{\rm GLS}(\hat{a}_s),
&=& 
1 -  \hat{a}_s 
+ \hat{a}_s^2 (-4.58333 + 0.33333 N_F)
+ \hat{a}_s^3 (-41.4399 + 7.74370 N_F - 0.17747 N_F^2) 
\nonumber\\ &&
+ \hat{a}_s^4 (-479.448 + 140.796 N_F - 8.39702 N_F^2 + 0.10374 N_F^3)~,
\end{eqnarray} 
%----------------------------------------------------------------------------------------------------------------------------- 
choosing the renormalization scale $\mu^2 = Q^2$, for $SU(3)_c$. Here $N_F$ denotes the number of 
active light flavors. The expression for general color factors was given in Ref.~\cite{Baikov:2012zn,Baikov:2010je}.
Note that the QCD corrections to the Gross-Llewellyn Smith sum rule and to the polarized Bjorken sum rule \cite{Bjorken:1969mm} 
are identical up to $O(\hat{a}_s^2)$.

For the asymptotic massive corrections (\ref{eq:WIL1}, \ref{eq:WIL2}) only the first moment of 
the massless Wilson coefficients 
$\hat{C}_{q,3}^{(2,3),\rm NS}(N_F)$ contributes, weighted by the first moments of the valence quark densities and
the corresponding CKM matrix elements, since the first moments of the massive non-singlet OMEs vanish due to fermion number 
conservation. This also holds at higher order. Unlike the case of the 
polarized Bjorken sum rule \cite{Behring:2015zaa}, in 
the present case two massive Wilson coefficients contribute. One obtains 
%---------------------------------------------------------------------------
\begin{eqnarray}
\label{eq:PBJ}
\int_0^1 {dx} \left[F_3^{\bar{\nu} p}(x,Q^2) + F_3^{{\nu} p}(x,Q^2)\right]
&=& 2 \Bigl[
  (|V_{du}|^2 +|V_{su}|^2) \langle u_v \rangle   
+ (|V_{dc}|^2 +|V_{du}|^2) \langle d_v \rangle   
\Bigr] 
\nonumber\\ 
&& \times C_{\rm GLS}(\hat{a}_s, N_F+1)
\\
&=& 
2 \left[2 \cdot 0.9999 + 0.9998\right]C_{\rm GLS}(\hat{a}_s, N_F+1)
\label{GLS1}
\end{eqnarray}
%-----------------------------------------------------------------------------------------------------------------------------
for the charm excitation with additional $N_F$ light flavors. Here we assumed $s_v = 0$ and 
$\langle q_v \rangle = \int_0^1 dx q_v(x)$ denotes
the first moment of a valence distribution. The CKM matrix elements currently imply a very small deviation from the factor 6 in 
(\ref{GLS1}) in the asymptotic case $Q^2 \gg m^2$ for proton targets. Very similar results are obtained for isoscalar targets.
There are, in particular, no logarithmic corrections in the mass scales involved. The heavy quark just induces a shift $N_F
\rightarrow N_F +1$ w.r.t. the case of $N_F$ light quarks only. Different results are obtained in the non-inclusive 
tagged-flavor case 
\cite{Buza:1997mg}, which has been calculated to $O(\alpha_s^2)$. Here no inclusive structure functions are 
considered unlike the case in Ref.~\cite{Blumlein:2014fqa}. Corresponding 
power corrections in the tagged-flavor case were derived in Refs.~\cite{Blumlein:1998sh,vanNeerven:1999ec}.

%%%%%%%%%%%%%%%%%%%%%%%%%%%%%%%%%%%%%%%%%%%%%%%%%%%%%%%%%%%%%%%%%%%%%%%%%%%%%%%%%%%%%%%%%%%%%%%%%%% 
\section{Conclusions} 
\label{sec:5} 
%%%%%%%%%%%%%%%%%%%%%%%%%%%%%%%%%%%%%%%%%%%%%%%%%%%%%%%%%%%%%%%%%%%%%%%%%%%%%%%%%%%%%%%%%%%%%%%%%%%
%---------------------------------------------------------------------------------------------------------------------

\vspace{1mm}
\noindent
We calculated the heavy flavor non-singlet Wilson coefficients for the charged current structure function 
$xF_3^{W^+-W^-}(x,Q^2)$ to $O(\alpha_s^3)$ in the asymptotic region $Q^2 \gg m^2$. They can be expressed
in terms of nested harmonic sums and polylogarithms. In contrast to the neutral current case, here, a
second heavy flavor Wilson coefficient contributes, which describes the flavor excitation due to a $s \rightarrow c$
transition. The heavy flavor contributions to this combination of structure functions 
amounts up to the $O(\pm 3 \%)$ for proton targets. Very similar results are obtained for deuteron targets,
as the present combination approximately selects isoscalarity. In the small and large $x$ regions the heavy flavor effects 
at 3-loop order are visible on top of those up to 2-loop orders by an enhancement and a depletion of $O(1\%)$, 
respectively, in the
ratio to the 3-loop massless Wilson coefficient. This is below the present resolution reached at HERA, but 
may become of importance in high luminosity measurements at future colliders. The scale dependence reached
for the non-singlet combination studied turns out to be widely flat. We also presented the leading terms 
in the heavy and light flavor Wilson coefficients at small and large values of the momentum fraction $x$.

In the asymptotic case $Q^2 \gg m^2$ the contribution of the massive Wilson coefficients to the Gross-Llewellyn Smith sum 
rule reduce to the replacement $N_F \rightarrow N_F + 1$ in the massless approximation. One has to note CKM-matrix 
effects here.

\appendix
%%%%%%%%%%%%%%%%%%%%%%%%%%%%%%%%%%%%%%%%%%%%%%%%%%%%%%%%%%%%%%%%%%%%%%%%%%%%%%%%%%%%%%%%%%%%%%%%%%% 
\section{Appendix} 
\label{sec:APP} 
%%%%%%%%%%%%%%%%%%%%%%%%%%%%%%%%%%%%%%%%%%%%%%%%%%%%%%%%%%%%%%%%%%%%%%%%%%%%%%%%%%%%%%%%%%%%%%%%%%%
%---------------------------------------------------------------------------------------------------------------------

\vspace{1mm}
\noindent
The massless Wilson coefficients and anomalous dimensions for $xF_3^{W^+}(x,Q^2)+ xF_3^{W^+}(x,Q^2)$
contain a new color factor proportional to $d^{abc} d^{abc}$ ($= 40/3$ in QCD) \cite{Moch:2008fj,NSANOM}.
This color factor does not appear in the massive operator matrix elements at
3-loop order, which we would like to discuss in the following. 
Those terms, however, do arise in individual diagrams contributing to
the OMEs but they cancel in the complete result. The cancellation can be seen as
follows.

\begin{figure}[H]
        \centering
        \includegraphics{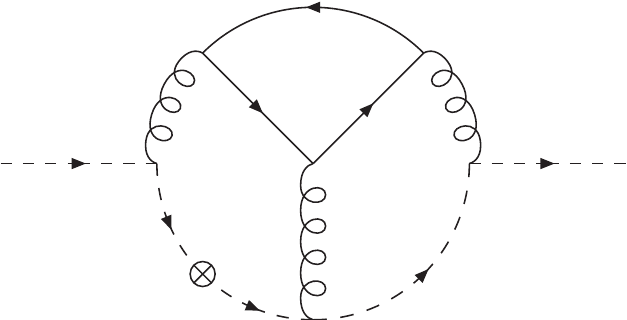}
        \hspace{1em}
        \includegraphics{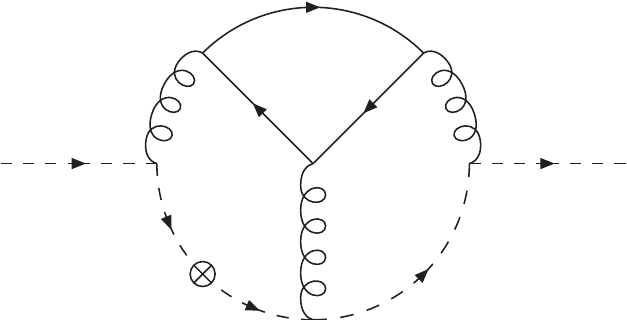}
        \caption{Example of a pair of diagrams which each contain a term
                 proportional to the color factor $d^{abc} d^{abc}$.}
        \label{fig:diagdabc}
\end{figure}

The color factor can appear when there are two separate fermion lines connected
by three gluons. One fermion line connected to three gluons yields the color
structure
%---------------------------------------------------------------------------------------------------------------------
\begin{eqnarray}
        \mathrm{Tr}[t^a t^b t^c] &=& \frac{T_F}{2} (i f^{abc} + d^{abc})~.
\end{eqnarray}
%---------------------------------------------------------------------------------------------------------------------
Thus, two fermion lines connected by three individual gluon propagators produce
the color structure
\begin{eqnarray}
        \mathrm{Tr}[t^a t^b t^c] \delta^{aa'} \delta^{bb'} \delta^{cc'}
        \mathrm{Tr}[t^{a'} t^{b'} t^{c'}] &=&
        \frac{T_F^2}{4} (-f^{abc}f^{abc} + d^{abc} d^{abc})~.
        \label{eq:dabctrace}
\end{eqnarray}
For each such diagram there is a corresponding diagram with the fermion flow
along the closed fermion loop reversed. An example of a pair of diagrams is
given in Figure~\ref{fig:diagdabc}. The closed loop has to have three quark-gluon
vertices in order to produce the color factor and therefore the loop must have
three fermion propagators. Keeping the direction of the momenta fixed, the
reversal of the fermion flow entails a change of the sign of the momentum
$\slashed{p}_i$ in the numerator of each fermion propagator,
%---------------------------------------------------------------------------------------------------------------------
\begin{eqnarray}
        \frac{i(\slashed{p}_i+m)}{p_i^2-m^2}
                &\to& \frac{i(-\slashed{p}_i+m)}{p_i^2-m^2}~.
\end{eqnarray}
%---------------------------------------------------------------------------------------------------------------------
%---------------------------------------------------------------------------------------------------------------------
%---------------------------------------------------------------------------------------------------------------------
This sign can be factored out since traces over an odd number of Dirac matrices
vanish and yields a global factor $(-1)$. Besides that, the reversal of the
fermion flow also reverses the order of the color generators $t^{a'} t^{b'}
t^{c'}$ in Eq.~\eqref{eq:dabctrace} which flips the sign in front of the
$f^{abc} f^{abc}$ term, but leaves the $d^{abc} d^{abc}$ term unchanged. We see
that each pair of such diagrams has exactly the same integrand, but the sign in
front of the $d^{abc} d^{abc}$ color factor is changed. Therefore, this color
factor cancels in the sum for each pair of diagrams.

\vspace{5mm}
\noindent
{\bf Acknowledgment.}~
We would like to thank J.~Ablinger, K.~Chetyrkin, M.~Round, A.~Vogt, and F.~Wi\ss{}brock for discussions. 
This work was supported in part by the European Commission through contract PITN-GA-2012-316704 ({HIGGSTOOLS}) 
and Austrian Science Fund (FWF) grant SFB F50 (F5009-N15). The Feynman diagrams were drawn using {\tt Axodraw}
\cite{Vermaseren:1994je}.

\newpage
%--------------------------------------------------------------------------------------------------

%----------------------------------------------------------------------------------------------
%------------------------------------------------------------------------------
\end{document}